\documentclass[aps,prl,twocolumn,superscriptaddress,showpacs,floatfix]{revtex4-1}

\usepackage{graphicx}
\usepackage{amssymb,amsmath}
\usepackage{color}
\usepackage{bm}
\usepackage{textcomp}

\begin{document}
%\title{Dominance versus alternation of states in the presence of aging}
\title{Competition in the presence of aging: order, disorder, and synchronized collective behavior}

\author{Toni P\'erez}
\affiliation{Instituto de F\'isica Interdisciplinar y Sistemas Complejos (IFISC), Spain}
\author{Konstantin Klemm}
\affiliation{School of Science and Technology, Nazarbayev University,
Kabanbay Batyr Ave.\ 53, 010000 Astana, Kazakhstan}
\affiliation{Bioinformatics, Institute of Computer Science, University
Leipzig, H{\"a}rtelstr. 16-18, 04107 Leipzig, Germany}
\affiliation{Bioinformatics and Computational Biology, University of Vienna,
W\"{a}hringerstra{\ss}e 29, 1090 Vienna, Austria}
\affiliation{Theoretical Chemistry, University of Vienna,
W\"{a}hringerstra{\ss}e 17, 1090 Vienna, Austria}
\author{V\'ictor M. Egu\'iluz}
\affiliation{Instituto de F\'isica Interdisciplinar y Sistemas Complejos (IFISC), Spain}

\begin{abstract}
% We approach the study of competition between states when aging of the states is present.
% First, we introduce a basic stochastic dynamics model with two states, up and down, and two ages, old and young. A mean-field scenario shows rich but analytically tractable dynamics. 
% When the population has a preference to adopt older states the system orders quickly due to the dominance of the old state. 
% When preference for new states prevails, the system can show coexistence of states or oscillations between the leading state resulting in long ordering times. 
% Finally, we explore the generality of the results by extending our study to a continuous age space obtaining a good agreement with the results of the basic model. 
% Implications for social systems are discussed. 
% For a basic model with two states, up and down, and two ages, old and young, a mean-field scenario shows rich but analytically tractable dynamics. 
% Varying the control parameter $\alpha$, the ordering time of the system is discontinous at $\alpha=0$ (corresponding to the unbiased voter model VM), and is strongly accelerated ($\alpha>0$) or 
% decelerated ($\alpha<0$) compared to VM. 
%Finally, we explore the generality of the results by extending our study to a continuous age space obtaining a good agreement with the results of the basic model.
We study the stochastic dynamics of coupled states with transition probabilities depending on local persistence, this is, the time since a state has changed. 
When the population has a preference to adopt older states the system orders quickly due to the dominance of the old state. 
When preference for new states prevails, the system can show coexistence of states or synchronized collective behavior resulting in long ordering times. 
In this case, the magnetization $m(t)$ of the system oscillates around $m(t)=0$. 
Implications for social systems are discussed.
\end{abstract}

% \pacs{89.20.-a}{Interdisciplinary applications of physics}
% \pacs{89.75.-k}{Complex systems}
% \pacs{89.65.-s}{Social and economic systems}

\maketitle

%
% INTRODUCTION
%

% The dynamics of adoption of traits such as innovations, opinions, 
% or ideas is a topic attracting attention of researchers from disciplines as diverse as Economics, Sociology, Biology and Physics 
% \cite{Rogers1962,Mahajan1985,Kleinberg2010,Young2011,Morris2000,Montanari2010,Caccioli2008,Sood2005,Castellano2009}.
Models of two states are commonly used in physics as a tool to study the emergence of collective behavior in systems from spin interaction to opinion dynamics \cite{Marro1999,SanMiguel2005,Castellano2009}. 
In the adoption of traits %like innovations, opinions, or ideas 
\cite{Rogers1962,Mahajan1985,Kleinberg2010,Young2011} 
different aspects have been studied including the relevance of the interaction topology \cite{Morris2000,Sood2005,Montanari2010}, 
social influence \cite{Latane1981,Moussaid2013}, and mass media \cite{Avella2007,Quattrociocchi2011,Hodas2014}. 
When accounting for opinion dynamics, the majority of models are based on decision rules that consider a fraction of the surrounding states, e.g., 
voter model \cite{Holley1975}, threshold model \cite{Granovetter1978}, majority rule \cite{Galam1986}, or Sznajd model \cite{Sznajd2000}. 
% However, when individuals make choices they usually relay on their own past experience or memory \cite{Miller2013,Rendell2010,Moinet2015}. 
The timing of the interactions can also affect the behavior of the system at least by two ways: the precise sequence of interactions and by the aging of states. 
For example, in epidemic spreading and diffusion, the temporal sequence of interactions can slow down the spreading process \cite{Karsai2011,Mieghem2013,Masuda2013,Wang2014}; 
in ordering dynamics, state-dependent updates can have a qualitative impact on the mean time to order \cite{Holme2012,Baxter2011,Stark2008,Caccioli2008,Fernandez2011,Takaguchi2011}. 
Aging in physical systems refers to the persistence time, that is, the time spent in a given state, and affects the response of the system to an external field or perturbation 
\cite{Young1998, Struik1978}.  
In social systems, when individuals make choices they usually relay on their own past experience or memory \cite{Miller2013,Rendell2010,Moinet2015}. 
While conservative groups tend to hold ideas in an unaltered form for a long time, progressive individuals embrace new opinions, ideas, or a technology and 
disseminate them with more enthusiasm \cite{Bass1969,Toole2012}. 
In the competition between new and old information, although new information is more valuable for exploring and spatial searching \cite{Lizana2010}, 
adopting older strategies can promote cooperation and group success \cite{Yang2014}. 
Also in a biological context, aging in speciation events has been proposed as a mechanism to explain the shape of evolutionary trees \cite{Keller2015}.
Here we analyze how the tendency of agents towards the adoption of established vs.\ novel traits influences the macroscopic dynamic and the ordering process.
We tackle this problem by considering a model in which the adoption of states depends on the time span the agents have held their current states.

%
% FUNDAMENTAL MODEL (four states)
%

\begin{figure}
\begin{center}
\includegraphics[width=0.9\columnwidth]{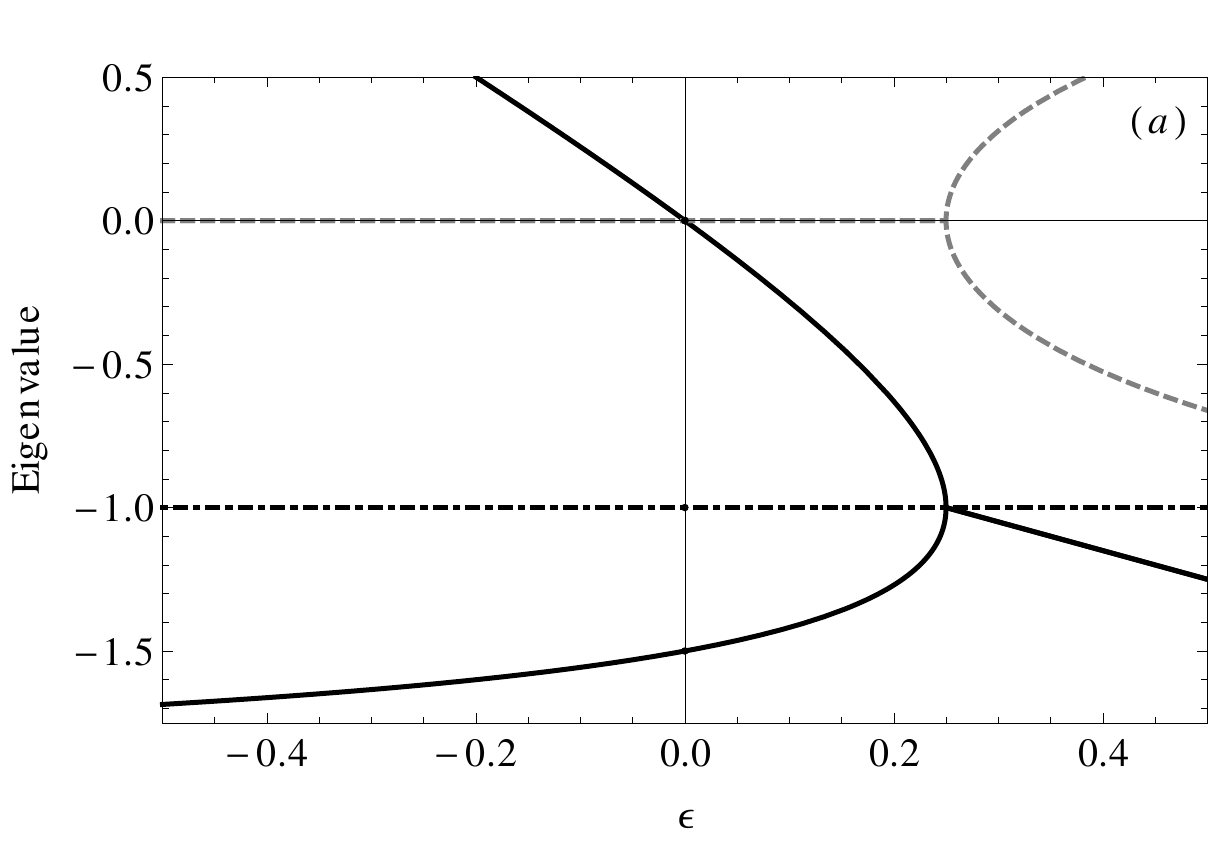}
\includegraphics[width=0.9\columnwidth]{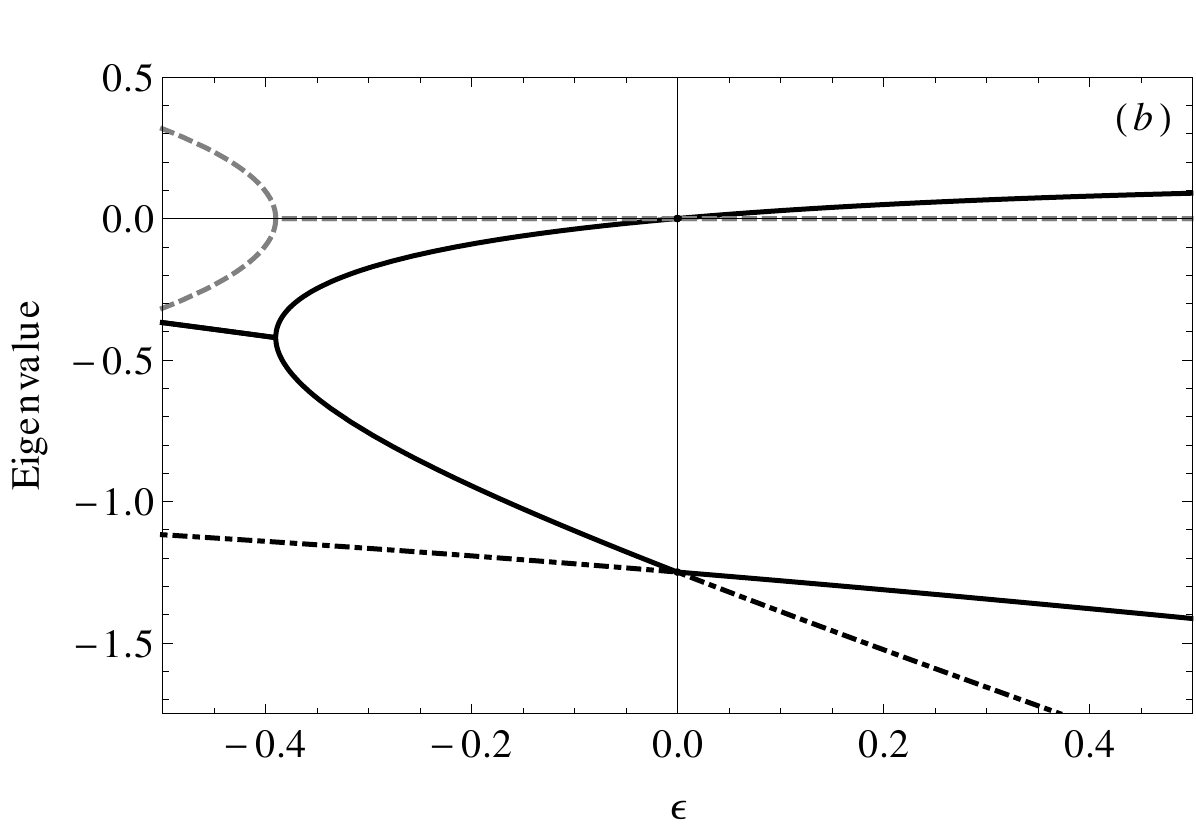}
\end{center}
\caption{\label{fig:eigen}
Eigenvalues of the stationary solutions of Eq.~(\ref{eq:sys4}) as a function of persuasiveness $\epsilon$ for: 
(a) the homogeneous solutions $S_1$ and $S_2$ having the same eigenvalues, and (b) the solution $S_3$. Black solid (gray dashed) lines represent real (imaginary) parts of the two complex conjugate eigenvalues. 
Doted-dashed black line represent the third eigenvalue (real). The fourth eigenvalue (not shown) for the eigenspace in $(1,1,1,1)$ direction is zero due to conservation of normalization.
}
\end{figure}

%\emph{Basic model.} 
The model is defined as follows: each agent has a state $l$ that can be up ($\uparrow$) or down
($\downarrow)$ with age young ($y$) or old ($z$). Agents can be then in four states $y^\uparrow$, $y^\downarrow$,
$z^\uparrow$, and $z^\downarrow$. Young agents become old at a rate that we set to $r=1$.
Then, there are reactions of randomly paired agents: i) in young $i$ and old $j$ of
opposite opinions, $i$ adopts the opinion of $j$ with probability  
$w =\frac{1}{2}+\epsilon$, otherwise (with probability $1-w$),
$j$ adopts the opinion of $i$; ii) in pairs of agents with the same age and different opinion, each agent has probability $\frac{1}{2}$ of convincing the other; 
iii) for pairs of agents with the same opinion, nothing happens. 
When an agent adopts an opinion, it goes to the young age of the adopted
opinion. Neglecting correlations, the expectation values of state concentrations evolve
according to
\begin{equation} \label{eq:sys4}
\begin{array}{lllllll}
\dot{y}^{\uparrow}   = & - y^\uparrow   + (1-2w) y^\uparrow z^\downarrow  + w y^\downarrow z^\uparrow  + \frac{1}{2} z^\uparrow z^\downarrow~,\\ \\
\dot{y}^{\downarrow} = & - y^\downarrow  + w y^\uparrow z^\downarrow  + (1-2w) y^\downarrow z^\uparrow  + \frac{1}{2} z^\uparrow z^\downarrow~,\\ \\
\dot{z}^{\uparrow}   = &  y^\uparrow    - (1-w) y^\downarrow z^\uparrow  - \frac{1}{2} z^\uparrow z^\downarrow~,\\ \\
\dot{z}^{\downarrow} = &  y^\downarrow  - (1-w) y^\uparrow z^\downarrow  - \frac{1}{2} z^\uparrow z^\downarrow~,
\end{array}
\end{equation}
with the normalization $y^\uparrow+ y^\downarrow + z^\uparrow+ z^\downarrow =1$. Here we use $y^{\uparrow\downarrow}$ and $z^{\uparrow\downarrow}$ to refer to 
the fraction of the corresponding states occupied by the agents. 
The parameter $\epsilon$ corresponds to the persuasiveness of the agent, $\epsilon >0$ means that agents with older opinions are more persuasive. 
On the contrary, $\epsilon <0$ corresponds to agents with younger opinions been more persuasive. 

%The stationary solutions can be computed by setting the temporal derivatives of Eqs.~(\ref{eq:sys4}) equal to zero. 

The system presents three stationary solutions in the relevant range of all four variables being non-negative. 
Two fixed points are the homogeneous solutions $S_1$ having $z^\downarrow=1$ and $S_2$ having $z^\uparrow=1$. Here either all opinions are down ($S_1$) or all are up ($S_2$) and old.
%and, since no interactions between different opinions occur, all opinions are old. We plot the Jacobian eigenvalues in dependence of $\epsilon$ in Figure~\ref{fig:eigen}(a), being
%the same for both homogeneous solutions. Concentrating on the principal eigenvalue, 
The homogeneous solutions are stable if $\epsilon >0$. Non-zero imaginary parts of two eigenvalues are obtained for $\epsilon>1/4$. 
The third fixed point $S_3$ is an up-down-symmetric solution with values $y^\uparrow=y^\downarrow=\frac{5+2\epsilon-\Delta}{8\epsilon}, z^\uparrow=z^\downarrow=\frac{-5+2\epsilon+\Delta}{8\epsilon}$ 
where $\Delta=\sqrt{25+4(\epsilon^2+3\epsilon)}$. As shown in Figure~\ref{fig:eigen}(b), it is stable if $\epsilon < 0$, thus complementary to the stability of the homogeneous
solutions. A transition from zero to non-zero imaginary parts of two eigenvalues occurs when $\epsilon$ falls below approximately $-0.39$. In this regime of strongly negative $\epsilon$, the system
oscillates when relaxing from a perturbation out of the symmetric fixed point solution $S_3$. 
This stability scenario is qualitatively maintained when $r$ changes. As $r\rightarrow0$, the point at which the non-zero imaginary part of the eigenvalues shows up shifts towards $\epsilon=0$.

%
% STOCHASTIC FINITE-SIZE MODEL
%
\emph{Model with continuous ages.} 
We now move from two ages to a continuous age space and introduce an age dependent probability. The model is now described as follows. Agents can be in one of the two opinions, up or down.  
The state of each agent has associated an age defined as: $\tau_i=t-t_i$, where $t$ is the current time and $t_i$ is the time 
when the current state of agent $i$ was acquired. The system evolves by randomly selecting a pair of agents that, if they are in different states, 
with probability $p_{i\rightarrow j}$ agent agent $j$ adopts the state of agent $i$, and with probability $1-p_{i\rightarrow j}$ the contrary happens $i$ copies the state of $j$. 
If the agents are already in the same state, no change is applied. After $N$ updates, time $t$ is increased to $t+1$.
We define the probability $p_{i\rightarrow j}$ to be dependent on the age of the states of both interacting agents as
%$p_i=\frac{\tau_i^{\alpha}}{\tau_i^{\alpha}+\tau_j^{\alpha}}.\label{eq.1}$
%$p_i=[1 + (\tau_j/\tau_i)^{\alpha}]^{-1}$.
\begin{equation}
 p_{i\rightarrow j}=\frac{1}{1 + (\tau_j/\tau_i)^{\alpha}}~.
\end{equation}

Initially, each agent has randomly assigned one of the two opinions and the initial ages $t_i$ are uniformly distributed proportionally to the system size $N$. %in $[0, 0.1N]$. % with $\Delta=0.1$. 
We consider random mixing where each agent is allowed to interact with any other agent. 
The case $\alpha=0$ corresponds to an updating probability of $p_{i\rightarrow j}=0.5$ which leads to voter model dynamics \cite{Holley1975}. 
Large values of the exponent ($\alpha \rightarrow \infty$), correspond to situations in which the agent with the initial oldest state 
is imposing her opinion to the others. In the other extreme case ($\alpha \rightarrow -\infty$), the youngest state is imposed. 
%%

%\emph{Absorbing configurations.}
For $\alpha \rightarrow \infty$ the system ends up in the state of the oldest opinion while for values of 
$\alpha~ \in~ (-\infty,0]$ the system adopts any of the two opinions with equal probability. For $\alpha~ \in~ (0,\infty)$ the probability that the system adopts the 
state of the initial oldest opinion grows with increasing $\alpha$ but it tends to $1/2$ when $N$ increases. 

\begin{figure}
\includegraphics[width=\columnwidth]{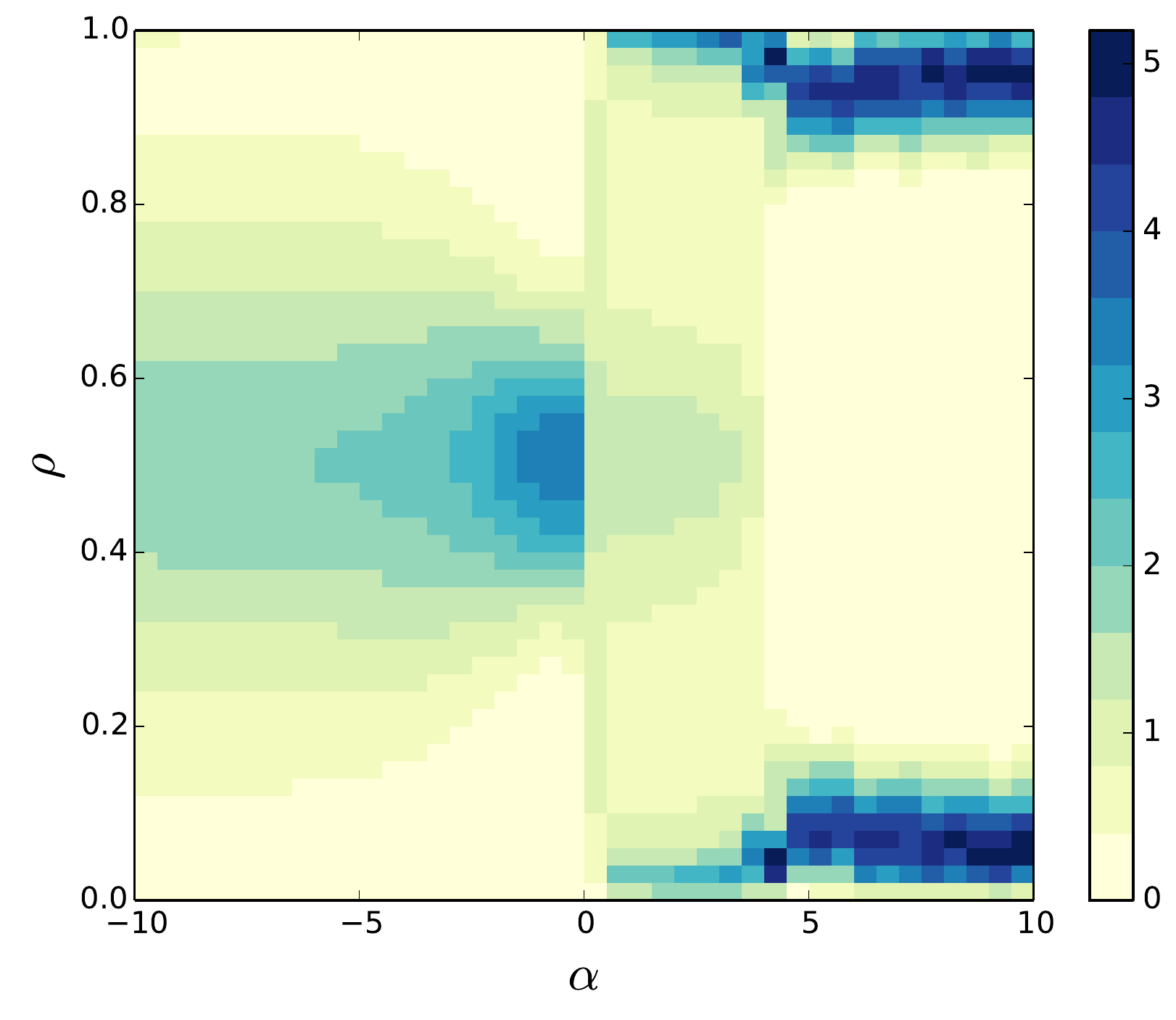}
\caption{Probability density function (codified as a colormap) of the dependence of the density of states with $\alpha$. The density of states is calculated by computing 
the normalized cumulative histogram of the fraction of population in each state during a fixed simulation period and averaged over $10^4$ realizations. The system size was fixed to $N=100$.}
\label{fig: 1.1.3}
\end{figure}

%\emph{Exploring the transition at $\alpha=0$.} 
%We observe a change in the behavior of $S_N(\alpha)$ when $\alpha$ crosses zero. 
Figure \ref{fig: 1.1.3} shows the probability distribution function of the density of states $\rho$ as a function of $\alpha$. For $\alpha=0$, the 
density of states is homogeneous corresponding to an equiprobable distribution of states. For $\alpha$ negative but close to zero, the dynamic is concentrated around $\rho=0.5$, 
which corresponds to a configuration where the agents alternate between any of the two opinions. This situation changes gradually to a more homogeneous distribution of states as $\alpha$ 
becomes more negative. For $\alpha>0$, the states are concentrated around $\rho=0$ and $\rho=1$ showing that the system eventually orders in one of the two 
opinions (the presence of the two peaks is due to the alternation in being the oldest opinion during the initial conditions). 

\begin{figure}
\includegraphics[width=\columnwidth]{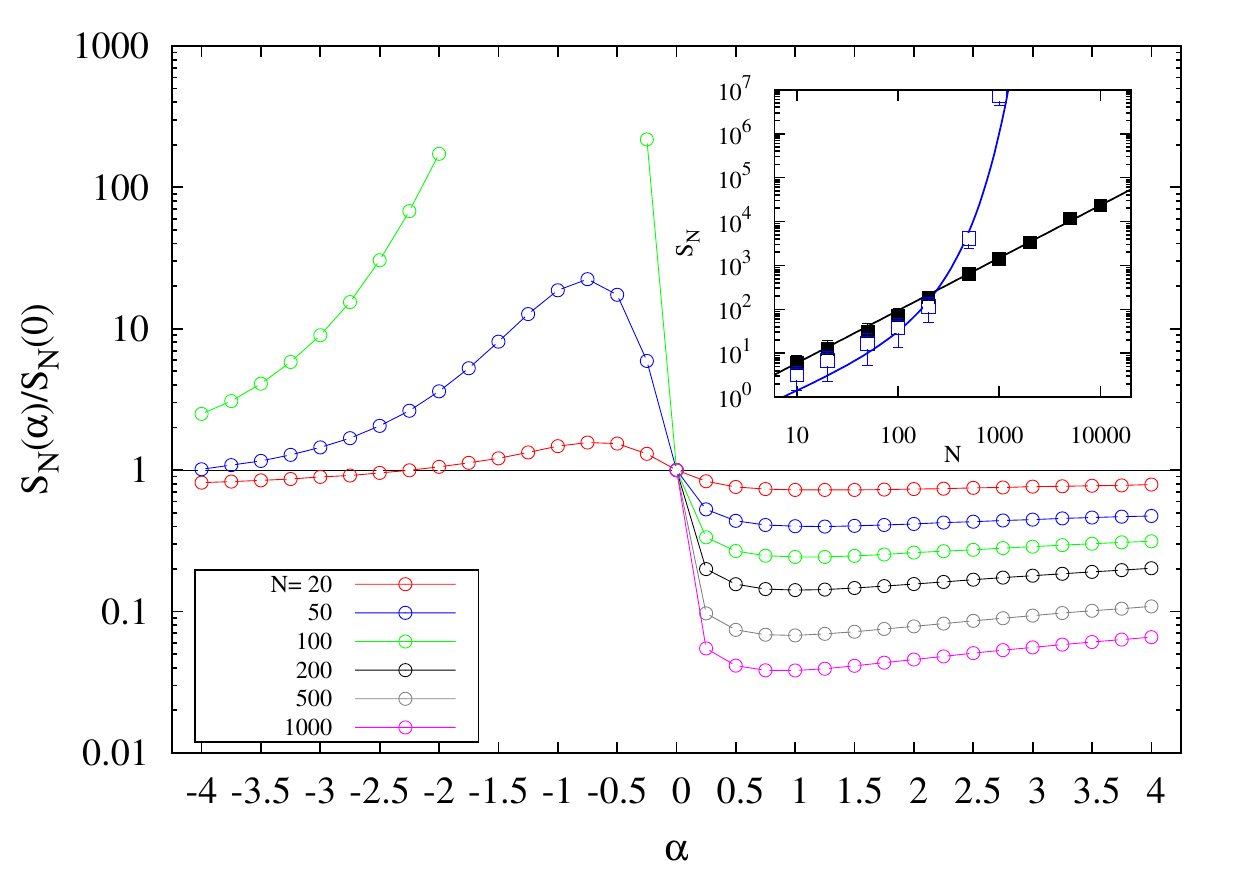}
\caption{Rescaled ordering time $S_{N}(\alpha)/S_{N}(0)$ versus $\alpha$ for different system sizes. 
Open symbols stand for the median of the ordering time normalized to the median of the ordering time at $\alpha=0$. 
The horizontal black line has been added for visualization purposes. Inset: Scaling of the median of $S_{N}(\alpha)$ 
in the limits $\alpha \rightarrow \infty$ (solid symbols) and $\alpha \rightarrow -\infty$ (open symbols). 
Dashed lines fit respectively $S_{N}(+\infty)\sim N^{\gamma}$ with $\gamma=1.2$ and $S_N(-\infty)\sim N\exp(bN)$ with $b=0.009$.}
\label{fig: 1.1.1}
\end{figure}

Figure \ref{fig: 1.1.1} shows the ordering time $S_{N}(\alpha)$, i.e., the time that the system needs to reach a final state where all the agents have the same opinion, 
computed as the median of the distribution of ordering times from different simulations and rescaled to the value $S_N(\alpha=0)$. %corresponding to the voter model ordering time.
$S_N(0)$ increases linearly with the $N$ as it does for the voter model \cite{Vazquez2008,Sood2005}.
For values $\alpha>0$, $S_{N}(\alpha)$ gets smaller than $S_{N}(0)$ implying that the system orders faster than in the voter model case. 
There is a transition when $\alpha$ crosses zero. For values $\alpha \lesssim 0$, $S_{N}(\alpha)$ increases very fast with $N$. This is in agreement with 
the observed dynamics around $\alpha=-1$ (see Fig.~\ref{fig: 1.1.3}). 
The extreme values of $\alpha$ correspond to the cases where, 
when confronting two states, the oldest opinion always induces the change ($\alpha=+\infty$) or the youngest opinion always induces the change ($\alpha=-\infty$). 
The inset of Fig \ref{fig: 1.1.1} shows the scaling with system size of $S_{N}(\alpha)$ in these limits. 
For $\alpha = +\infty$ the ordering time scales as $S_{N} \sim N^{\gamma}$ with the exponent $\gamma=1.2$. 
In the other limit, for $\alpha = -\infty$, the ordering time scales as $S_{N} \sim N \exp(bN)$ with $b=0.009$. 

\begin{figure}
\includegraphics[width=\columnwidth]{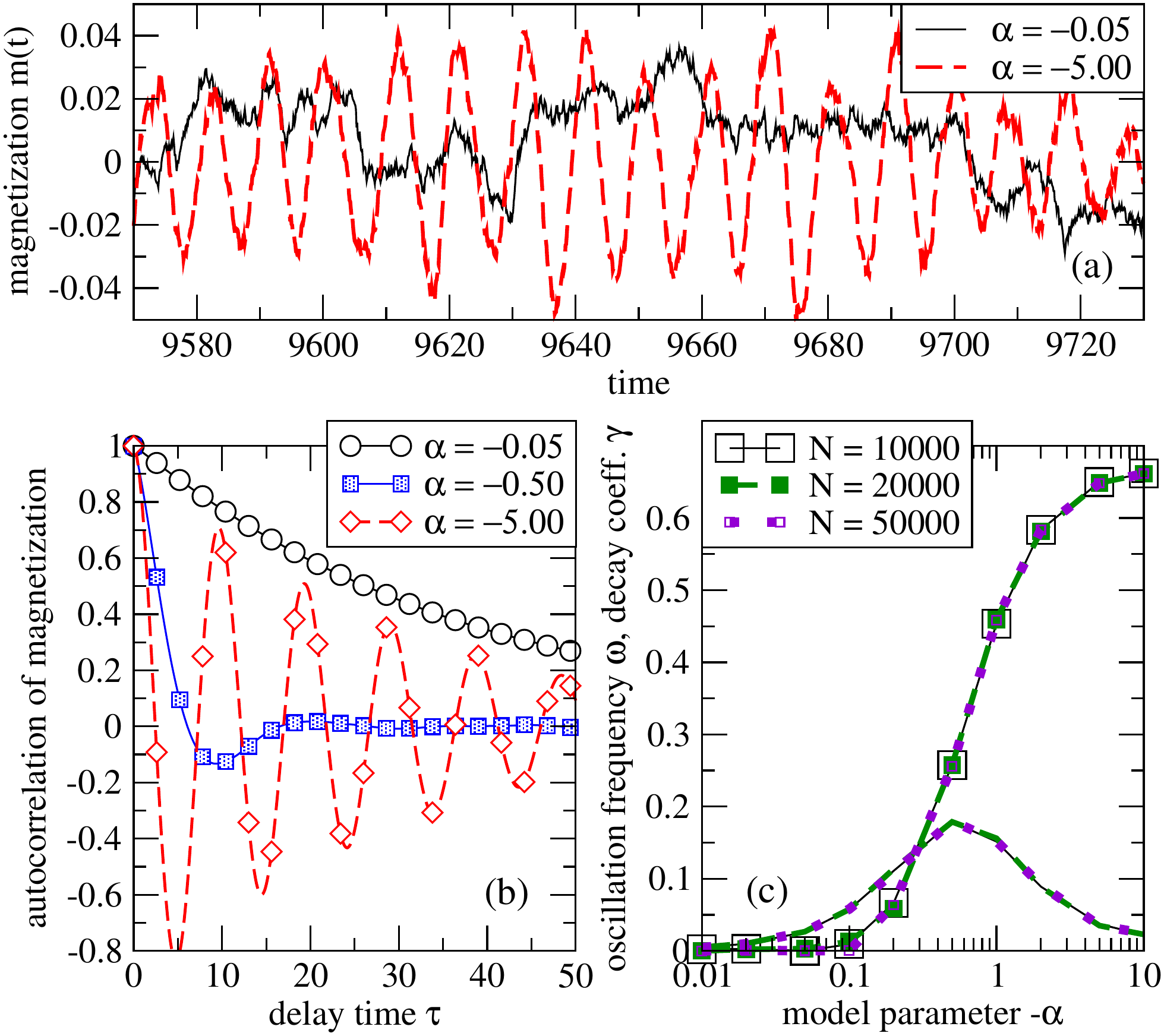}
\caption{\label{fig:osn}
Oscillations and decay of correlations for the finite-size model in the regime $\alpha<0$.
(a) Excerpts from time series of the magnetization for two different values of $\alpha$, system size $N=50000$. 
(b) Autocorrelation functions from magnetization time series of length $T=10^5$, system size $N=50000$. 
(c) Oscillation frequency $\omega$ (curves with symbols) and decay coefficients $\gamma$ (no symbols)
extracted from time series under different values of $\alpha$ and $N$. Curves for different system sizes $N$ are almost indistinguishable. 
Autocorrelation functions $A(\tau)$, $\tau\in[0,100]$, are considered for a least squares fit of the functional form $A_\text{fit}(t) = \exp(-\gamma t) \cos(\omega t)$.
}
\end{figure}

%Oscillations
In the regime $\alpha<0$, what is the behaviour of the system during the long ordering
times? Figure~\ref{fig:osn}(a) shows time series of the magnetization of the system. For
$\alpha$ negative and sufficiently far from zero, the magnetization oscillates around zero. 
Figure~\ref{fig:osn}(b) provides further analysis by the
autocorrelation functions of the magnetization time series. The onset of oscillations
is observed when $\alpha$ passes a value around $-0.5$ from above.
Figure~\ref{fig:osn}(c) shows frequency $\omega$ and decay constant $\gamma$ extracted
from the autocorrelation functions. These values do not exhibit significant dependence
on system size. The decay constant is maximum at the onset of oscillations, i.e.\ where
the frequency $\omega$ becomes non-zero. Both the onset of oscillations and the decay
behaviour are captured by the basic model, cf.~Figure~\ref{fig:eigen}(b). At the
transition to non-zero imaginary parts (oscillations), the stability of the symmetric
fixed point solution ($S_3$) is maximal, meaning that perturbations decay fastest.

To understand further the dynamic around $\alpha=0$ we define a quantity called the convincingness $z$. 
Let $S^+, S^- \in [N]$ be the two sets of agents with equal opinion within
each set and different opinions across sets. We define the {\em convincingness}
of $S^+$ vs.\ $S^-$ as the probability $z$ that the interaction of a uniformly
random pair of an $S^+$ agent and an $S^-$ agent leads to adoption of the
$S^+$ opinion, 
\begin{equation}
 z = |S^+|^{-1} |S^-|^{-1}  \sum_{i \in S^+} \sum_{j \in S^-}
 \frac{\tau_i^\alpha}{\tau_i^\alpha + \tau_j^\alpha}. 
%[1 + (\tau_j/\tau_i)^{\alpha}]^{-1}
\end{equation}
In case $\alpha<0$, there are competing effects governing the dynamics
of $z$. When $i \in S^+$ convinces $j \in S^-$, i) the set $S^+$ gains another member who now has the youngest opinion increasing $z$. 
ii) The set $S^-$ loses a member $j$ with $\tau_j$ typically larger than
average, making opinions of $S^-$ members younger on average decreasing
$z$. iii) With time advancing, all opinions age by the same additive rate.
This makes ratios between ages smaller, driving $z$ towards $1/2$. 
In the case $\alpha \ll -1 $, the first effect dominates. Thus, an
initial advantage in $z$ is amplified and the system orders quickly. 
For $\alpha \approx -1$, ordering times are large due to 
dominance of the second and third effects. In order to verify this idea,
we numerically record data pairs $(z(t),z(t+\tau)-z(t))$ with $\tau=1$. 
Averaging over pairs with the same or similar $z(t)$ values, 
we obtain $\langle z(t+\tau)-z(t) \rangle$ as the expected restoring force. 
The corresponding standard deviation is the noise strength at this $z$ value. 
The restoring force for $z$ is linear around the equilibrium point $z(t)=0.5$ while the noise
strength is mostly independent of $z$ (see Fig.~1 at Supplemental Material). This suggests to picture the dynamics
around $\alpha=0$ as one-dimensional equilibrium in a hyperbolic potential under state-independent
additive noise.

\begin{figure}
\includegraphics[width=\columnwidth]{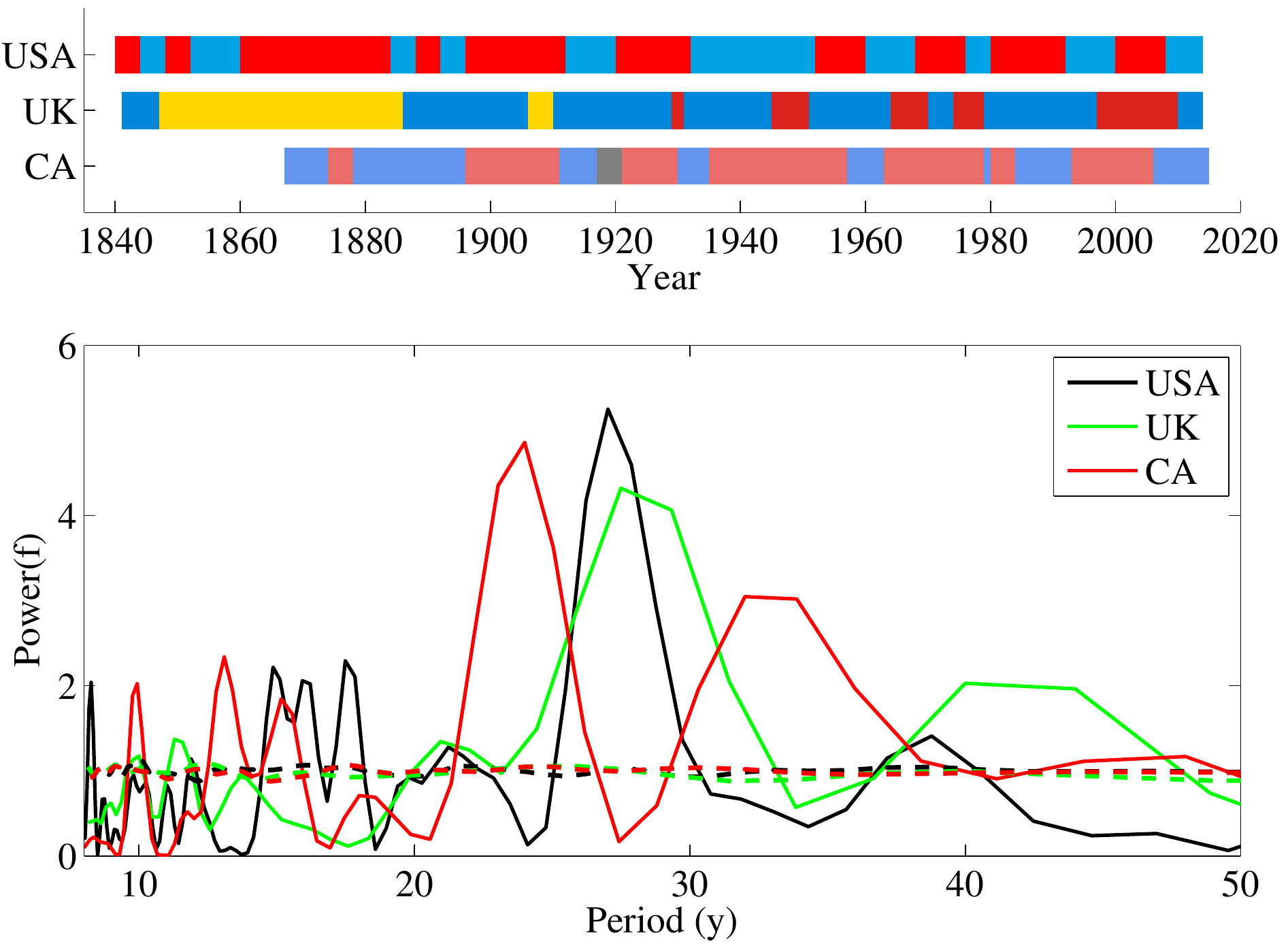}
\caption{\label{fig:elections}
Election results for United States, United Kingdom, and Canada. Elected parties are represented by their official colors, USA: Republicans (red) and Democrats (blue); 
UK: Conservative (blue), Liberals (yellow), and Labour (red); Canada: Liberals (red), Unionist (gray), and Conservatives (blue).
Bottom panel: Lomb periodograms of the binary time series for each country. Dashed lines represents the level of noise as the result of shuffling the data $250$ times.
}
\end{figure}

Different real systems display dominance such as in the adoption of innovations \cite{Rogers1962,Mahajan1985} and alternation as in opinion formation dynamics \cite{Mayer1992,Roemer1995,Soubeyran2008} 
or economic cycles \cite{Gualdi2015}. 
As an example, Fig. \ref{fig:elections} shows the electoral results of the governmental elections for United States, United Kingdom, and Canada during several decades \cite{Elec_data}. 
The Lomb periodogram \cite{Lomb1976} of the binary time series reveals the existence of alternation between the political parties, by the presence of prominent peaks well above the noise level 
(shuffling of the data), with periods of $20-30$ years in agreement with observations \cite{Goertzel2005}. 
This period of time coincides approximately with the length of a generation. %It is frequently observed that young people are more receptive to emerging cultural trends. 
Different mechanisms have been proposed to explain political cycles: electorate disappointment \cite{Schlesinger1986}, voters mood changes \cite{Klinberg1952,Goertzel2005}, 
negativity effect \cite{Aragones1997} or policy inertia \cite{Soubeyran2006}. Our study complements those mechanisms and contributes to understand how a 
continuous age state dynamics model with competition between preference for old versus young opinions leads to alternation in the leading opinion.

Summarizing, we have studied the competition of states using a basic model that takes into account the aging of the state. 
The stability analysis of the solutions reveals the existence of two stable solutions for positive values of the persuasiveness (old opinion prevails) that compete for consensus. 
For large negative persuasiveness (young opinion prevails), only one solution is stable leading to oscillatory transients. 
We have extended our study to a more detailed continuous age model finding that, when confronting two opinions, 
the final configuration where only one opinion survives and the time needed to reach it is noticeably 
sensitive to the age of the state through the exponent $\alpha$ of the convincing probability. 
% In the limit $\alpha=\infty$ (oldest opinion is imposed), the ordering time scales as $S_{N}= cN^{\gamma}$ with an exponent $\gamma=1.2$. 
% Whereas for $\alpha=-\infty$, the consensus is formed by the younger feature in a time that goes as $S_N=N\exp(N)$. 
% For $\alpha \lesssim 0$ a transition in $S_N(\alpha)$ is observed where the ordering times diverge. 
%Studying the convincingness of one of the opinions in this region suggests a dynamic scenario as in one-dimensional equilibrium in a hyperbolic potential under state-independent additive noise. 
The continuous age model exhibits likewise the oscillations shown by the basic model for large negative persuasiveness as well as similar solutions with $\alpha$. 
Our study provides an alternative mechanism in the understanding of the dynamics of consensus formation and the observed alternation between states of different systems.

\acknowledgments
The authors acknowledge support from project MODASS (FIS2011-24785) and Volkswagen Foundation.
TP acknowledges support from the program Juan de la Cierva of the Spanish Ministry of Economy and Competitiveness.

\end{document}